**Commentary on the Interpretation of Fourier-Transform Scanning Tunneling Microscopy Data**

Chris Mann, Nanohmics, Inc., Austin, TX 78741

Fourier transform scanning tunneling microscopy (FT-STM) is a powerful technique for investigating quasiparticle excitations on the surface of a material, providing access to momentum-space information from real-space data.[1, 2]  Near defects, intentionally introduced or innate, standing electronic waves can be produced that reveal scattering vectors, scattering selection rules, and the surface band structure through Fourier analysis of quasiparticle interference (QPI). Throughout the last two decades, FT-STM has been successfully applied to a variety of systems including 2D electron gases and high temperature superconductors.[3, 4]  In most of these systems, the scattering centers were essentially point defects on the surface; at a point defect, the incoming quasiparticle states may elastically scatter, creating an interference pattern with wavevector $\mathbf{q}=\mathbf{k}_{initial}-\mathbf{k}_{final}$. To access this scattering vector information, STM researchers produce a voltage-dependent map that is proportional to surface density of states by mapping the derivative of the current with respect to voltage (a dI/dV image), typically using lock-in techniques. The dI/dV map is then Fourier-transformed, resulting in a map that may be proportional to the $\mathbf{q}$ scattering vectors.

The scattering vectors that are observed, and their relative magnitude, is dependent on a number of factors: which states are available, which scattering channels are allowed, the coherence length of the state, and whether they are scattering to and from stationary states in the band structure, for instance. However, there are other subtle considerations that have been overlooked in many recent papers that can greatly influence the data and its interpretation.  An underlying principle of FT-STM is that each scatterer is much smaller than the wavelength and is circularly symmetric—in such a case, the incident and scattered waves will be reinforced enough to become STM-observable standing waves only if there are stationary states to scatter to and from.  Significantly, the shape of the scatterer has fundamental and often overlooked consequences for the interpretation of the data.

If the defects are *not* small relative to the wavelength, other concerns arise: (1) a soft or extended scatterer can broaden the incoming and outgoing wave phases, and (2) a circular wave scattering from a non-circular defect can produce a non-circular standing wave.  For Case 1, the shape of the potential can completely change the nature of the scattering event; a contrived example is the case of a reflectionless potential, where the shape of the potential, e.g., sech²(x), causes the incident wave's phases to conspire in a manner that prevents backscattering entirely and only requires a simple isotropic wave of the $e^{ikx}$ class, not a topologically non-trivial helical Dirac fermion.[5]  For Case 2, the Born approximation makes it clear that an isotropic wave encountering an anisotropic scatterer will result in anisotropic scattering: the scattered wave, $\psi_s$, is governed by the quasiparticle Green's function, $G(r)$, and the scattering potential, $V(r_0)$ through the integral equation $\psi_s(r) = \psi_{inc}(r) + \int G(r-r_0)V(r_0)\psi_s(r_0)d^3r_0$. On the surface of a crystal, the incident waves, $\psi_{inc}$, are assumed to be generated isotropically.  If $V$ is anisotropic, $\psi_s$ will likely be anisotropic as well, regardless of the symmetry of $G$. Clearly, the quasiparticle's Green function and the scattering potential are intrinsically linked and cannot be separated without including quantitative knowledge of one or the other.  This step is often skipped, and the data is analyzed without considering the geometry or potential of the scatterer.

Both of these cases, soft scatterers and non-circular scatterers, are present in many new materials systems including $Bi_2Se_3$ and $Bi_2Te_3$ where the predominant defects are diffuse triangles[6] and the obtained FT-STM data is hexagonal (which is the Fourier transform of a triangular feature).  The situation is complicated by the fact that, in systems with anisotropic scattering potentials, the band structure itself is already known to warp the surface state[7]—because the origin of the surface state anisotropy and the anisotropic potentials are innately linked through the crystal and band structure, deconvoluting and interpreting the observed features is not trivial. For instance, the claim that this illustrates hexagonal warping of the surface state is appealing, but could just as easily be explained by a traditional, circularly-symmetric, band-bending-induced 2DEG[8] scattering from triangular defects.

Another recent trend is the artificial symmetrizing of FT-STM data without discussing the process in the papers or supplementary sections—while this is a straight-forward image processing technique performed to reduce noise, there are implicit assumptions being made that are not discussed. To briefly overview the process: the data is multiplied by a window function, as is common for FFT analysis in most fields, in order to reduce the edge effects created by the finite image size, and also acts as smoothing filter of the FFT data (multiplication by a window in the real domain is a convolution in the Fourier domain). The magnitude of the FFT is calculated and the data is registered such that the zero-frequency peak is centered in the image. As the FFT is of real (non-imaginary) values, the FFT is guaranteed to be inversion symmetric about the center. Here the question arises: is it valid to take advantage of known structural symmetries of the crystal to reduce the noise in the data?

It is immediately obvious that symmetrization is not valid for systems with strong piezo hysteresis, thermal drift, or possible miscalibration, but more subtle considerations suggest that the technique is more fundamentally flawed; in fact, the entire concept behind the symmetrization of FT-STM data appears to be without literature support. Specifically, what is overlooked is that the positions of the defects themselves will necessarily be part of the Fourier transform and impact how the complex phases add up, ultimately influencing the structure of the plot.

To explicitly illustrate the concern, consider a set of randomly-distributed scatterers and the following wave model, such that each scatterer contributes to the real-space density of states as $f(\vec{r},\vec{r_0}) = A\cos(k|\vec{r}-\vec{r_0}|)/(\epsilon+|\vec{r}-\vec{r_0}|)^\gamma$, where $r_0$ is the position of a defect, $\epsilon$ is a broadening parameter to remove the singularity at the defect, $\gamma \geq 1$ is the falloff parameter to account for a finite coherence length and a geometric intensity falloff, and secondary and higher-order reflections are neglected. For raw FFT image $M$, the symmetry is enforced through the following operation:

$$M' = (M + R[M])/2$$

$$M'' = (M' + \theta\left[M',\frac{\pi}{3}\right] + \theta\left[M',\frac{2\pi}{3}\right])/3$$

$\theta[\cdot,\varphi]$ represents the rotation operation of angle $\varphi$ and $R[\cdot]$ represents the reflection operation. Some publications do not perform the first symmetrization step, which results in a chiral-looking FFT structure. Several simulations produced by this model are shown in Figure 1. In each of the three shown cases, the same circularly-symmetric wave functions are scattered off a different distribution of defects. It's clear that symmetrizing the data results in impressive images with high-contrast features, but the structure from one data set to the next is different—the number of scatterers and the scattering parameters are identical, so the only difference is the scatterer distribution. For instance, the circular structure appears to have a hexagonal warping, additional rings may be present that are not meaningful, and non-electronic noise appears as features.

The ability to reduce noise in an FFT is not limited to symmetrization: an alternative method of noise reduction is illustrated in Figure 2, where averaging of the FFT data from multiple ensembles of scatterers reduces the noise in a less-biasing manner. It's important to note that even when 9 simulated FFTs are averaged, their symmetrized data can still include artificial structure. The symmetrization process certainly reduces noise, but that's also part of the problem: the noise it leaves behind can look like a well-defined feature, leading to over-interpretation of data and making a poor alternative to traditional data averaging.

In cases with extended defects or asymmetric defects, found in a variety of important materials, the Fourier transform will reveal just as much about the scattering centers as it will about the scattered waves. The shapes of the defects are typically dictated by the structural nature of the crystal and can systematically introduce bias to the FFT data in a way that averaging cannot account for. Figure 3 illustrates how severe this effect can be through a series of decreasing wavelengths. For long wavelengths, Figure 3 top, the size of the defect is comparable to the wavelength and the dimensional effects are minimized (at least, for non-symmetrized data). For short wavelengths, Figure 3 bottom, the size of the defect is much larger than the wavelength, resulting in hexagonal warping of the FFT. The scattering centers must be as thoroughly characterized as the scattered waves to be aware of potential systematic bias that fundamentally alters the interpretation of the data.

Two last artifacts worth noting are from the defects themselves. In some cases, if the defect has a localized impurity state, it can produce a dI/dV signature on the surface that is driven by local bonding, rather than the band structure.[9] Typically, these signatures are non-dispersive in energy, which is one means of distinguishing an impurity state.[10] Another feature that may influence the dI/dV signatures in a scan is caused by extended physical protrusions or depressions on the surface; the STM data will demonstrate what appears to be an electronic state, but could just as easily be a variation in the apparent barrier height.[11] These artifacts will influence the shape of standing waves by biasing the center of the wave and can result in inaccurate FT-STM data, particularly for large defect densities.

It should be noted that there are other approaches available within the field of STM to extract k-space information: when investigating quasiparticles scattered by linear atomic steps, the geometry of the scattering potential is simple to account for—therefore, the concerns raised here do not affect the results from any of these studies. And in situations with long coherence lengths and few scatterers, the underlying electronic structure will have higher fidelity, as shown in Figure 4—this is essentially the difference between the near-field and far-field effects, where local effects caused by the anisotropy of the scatterers can decay faster than the long-range propagation. However, there are still influences from the near-field effects. Note that a larger dI/dV image will not provide the equivalent of a longer persistence length.

All of these concerns do not categorically contradict any existing data or claims, but they show that the correct interpretation of this data is far more complicated than previously treated and requires more detailed investigations to support the claims that have been made.

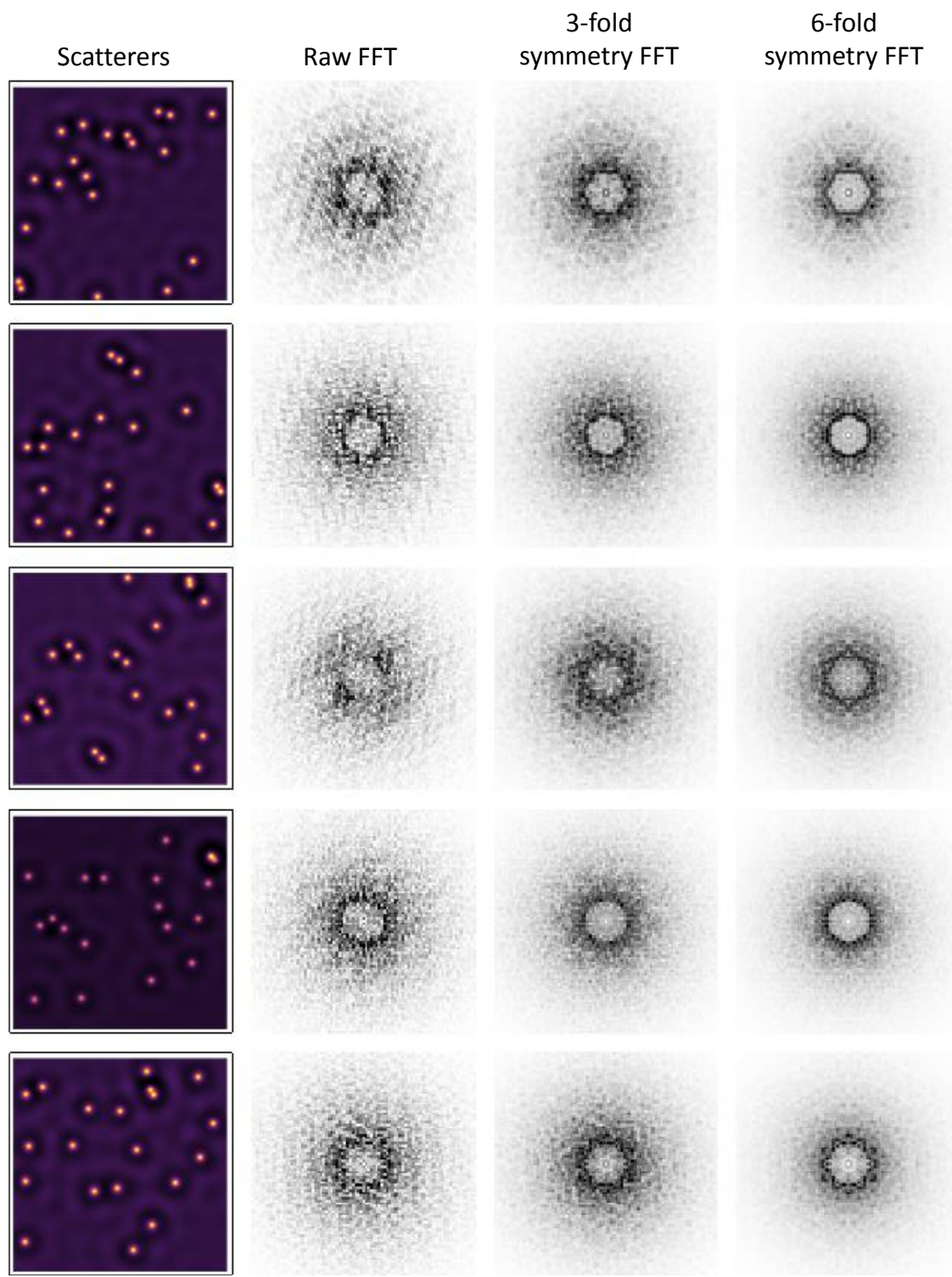

**Figure 1.** Several examples of artificial STM data based on identical parameters but different defect distributions (top row), the magnitude of the raw FFT, and the result of enforcing 3- and 6-fold symmetry. In all cases, the simulated waves have circular symmetry. When the symmetry is enforced, however, noise does not get cancelled out symmetrically, creating artificial structure in the symmetrized data including satellite peaks, chirality, additional rings, and apparent hexagonal warping. $k = 0.2$, $\epsilon = 1$, $\gamma = 7/4$, and A = 1.

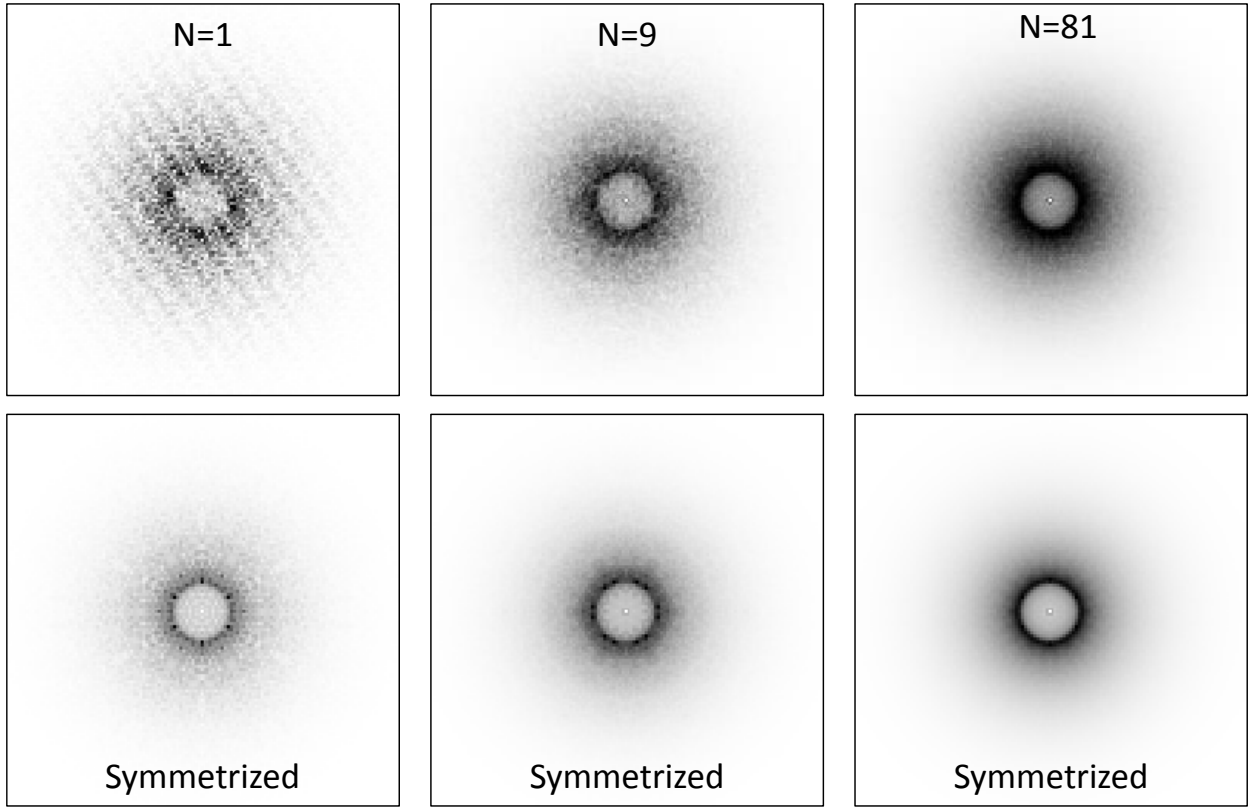

**Figure 2.** Averaged FFT data from N=1, 9, and 81 ensembles, each with 20 scatterers. Clearly, the 6-fold symmetrization process reduces the noise, but it can also produce structure in the data. Averaging multiple FFTs provides a compelling method of noise reduction, but it can be challenging to obtain enough data to perform this operation. Symmetrization of the average of 9 samples still results in some interesting, and meaningless, features such as high-intensity points within the ring structure. Symmetrization of the average of 81 samples produces the highest quality data, but as described later, even this may fail to produce meaningful information if the scattering centers are not circularly symmetric. $\epsilon = 1, \gamma = 7/4$, and A = 1.

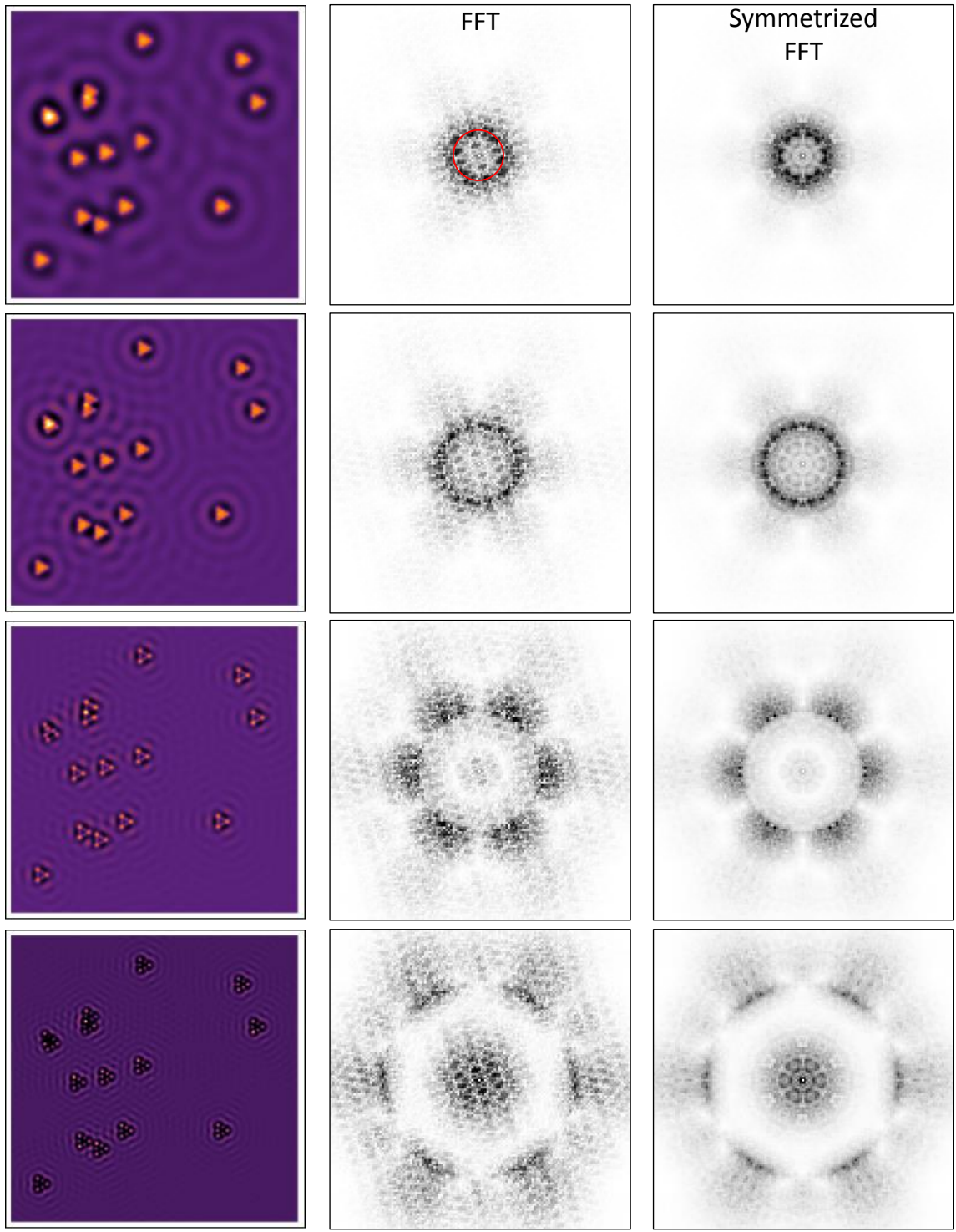

**Figure 3.** Results for different *k* values scattering from the same distribution of triangular scatterers, simulated using six equally-spaced point sources. For long wavelengths, the influence of the triangular structure is negligible, but when the wavelength approaches the physical extent of the defects, hexagonal 'warping' is apparent. In all of these cases, the symmetrization process exacerbates the problem and does not reveal the fact that the 'true' wavefunction is fully circularly symmetric. Averaging over multiple datasets will not help, either: rigorous analysis of the FFT will not reveal the truly circular nature of the waves. $\epsilon = 1$, $\gamma = 7/4$, and $A = 1/6$.

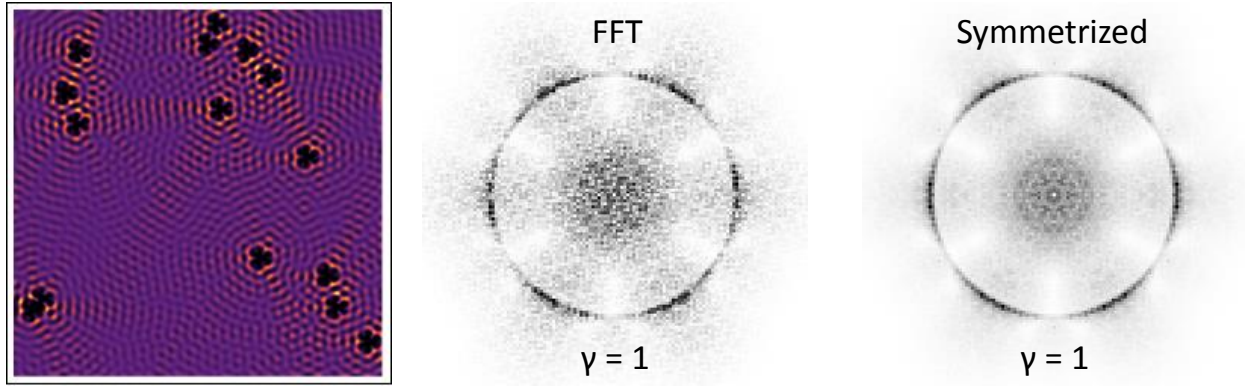

**Figure 4.** Longer coherence lengths results in improved data from the FFT ($\gamma = 1$), but do not result in perfect recovery of the circularly-symmetric wavefunction. Shorter coherence lengths (previous figure) are more strongly impacted by the near-field effects.

1. Petersen, L., et al., *Direct imaging of the two-dimensional Fermi contour: Fourier-transform STM.* Physical Review B, 1998. **57**(12): p. R6859.
2. Petersen, L., et al., *Fourier Transform-STM: determining the surface Fermi contour.* Journal of Electron Spectroscopy and Related Phenomena, 2000. **109**: p. 97-115.
3. Hoffman, J.E., et al., *Imaging Quasiparticle Interference in Bi2Sr2CaCu2O8+δ.* Science, 2002. **297**: p. 1148-1151.
4. Petersen, L., et al., *Screening waves from steps and defects on Cu(111) and Au(111) imaged with STM: Contributions from bulk electrons.* Physical Review B, 1998. **58**(11): p. 7361-7366.
5. Crandall, R.E. and B.R. Litt, *Reassembly and Time Advance in Reflectionless Scattering.* Annals of Physics, 1983. **146**: p. 458-469.
6. Urazhdin, S., et al., *Surface effects in layered semiconductors Bi2Se3 and Bi2Te3.* Physical Review B, 2004. **69**(8).
7. Hasan, M.Z. and C.L. Kane, *Colloquium: Topological insulators.* Reviews of Modern Physics, 2010. **82**(4): p. 3045-3067.
8. Bianchi, M., et al., *Coexistence of the topological state and a two-dimensional electron gas on the surface of Bi(2)Se(3).* Nat Commun, 2010. **1**: p. 128.
9. Urazhdin, S., et al., *Scanning tunneling microscopy of defect states in the semiconductor Bi2Se3.* Physical Review B, 2002. **66**(16).
10. Mann, C., et al., *Mapping the 3D surface potential in Bi(2)Se(3).* Nat Commun, 2013. **4**: p. 2277.
11. Hamers, R.J., *Atomic-resolution surface spectroscopy with the scanning tunneling microscope.* Annu. Rev. Phys. Chem., 1989. **40**: p. 531-559.